\RequirePackage{fix-cm}
\documentclass{svjour3}                     
\smartqed  
\usepackage{graphicx}
\usepackage{placeins}
\begin{document}

\title{A method to align time series segments based on envelope features as anchor points
}


\author{Cecilia Jarne         \and
        Pablo N. Alcain 
}


\institute{Cecilia Jarne \at
              Universidad Nacional de Quilmes - Departamento de Ciencia y Tecnolog\'ia CONICET
              \email{cecilia.jarne@unq.ed.ar}           
           \and
           Pablo N. Alcain \at
              Departamento de Física FCEN, UBA
}

\date{Received: date / Accepted: date}

\maketitle

\begin{abstract}
In the time series analysis field, there is not a unique recipe for studying signal similarities. On the other hand, averaging signals of the same nature is an essential tool in the analysis of different kinds of data. Here we propose a method to align and average segments of time series with similar patterns. A simple implementation based on \textit{python} code is provided for the procedure. The analysis was inspired by the study of canary sound syllables, but it is possible to apply it in semi periodic signals of different nature and not necessarily related to sounds.

\keywords{signal alignment and time series and averaging}
\end{abstract}

\section{Introduction: a general context regarding time series}

As defined in previous works, time series is a collection of observations made chronologically. The nature of time series includes: large in data size, high dimensionality and necessary to update continuously \cite{Fu,Esling,signals_book}. 

Many analyses in experimental research rely on the study and treatment of time series. The tasks regarding the analysis mainly can be categorized into: representation and indexing, similarity measures, segmentation, visualization, and mining.

In the context of time series data mining, a fundamental problem is how to represent the time series data. Different mining and searching tasks can be found in the literature and they can be roughly classified into the following fields: pattern discovery and clustering, classification, rule discovery and summarization \cite{Fu}.

Regarding the search for regularities, the partial periodic patterns are an important class existing in a time series analysis. The regularities can be classified into two types: (i) regular patterns: patterns exhibiting periodic behavior throughout a series with some exceptions and (ii) recurring patterns: patterns exhibiting periodic behavior only for particular time intervals within a series \cite{Kiran2015DiscoveringRP}.


Time-series mining task also requires a notion of similarity between series, based on the more intuitive notion of shape. A similarity measure should be consistent and provide the following properties:

\begin{itemize}
\item Provide a recognition of perceptually similar objects, even though they are not mathematically identical.
\item Be consistent with human intuition.
\item Emphasize the most salient features on both local and global scales.
\item Be universal in the sense that it allows to identify or
distinguish arbitrary objects, that is, no restrictions on time series are assumed.
\item Abstract from distortions and be invariant to a set of transformations.
\end{itemize}

Similarity measures can be classified into four categories. First, there are shape-based distances. They compare the overall shape of the series. Second, edit-based distances compare two time series on the basis of the minimum number of operations needed to transform one series into another one. The third category is Feature-based distances. These include extracting features describing aspects of the series that then are compared with any kind of distance function. The last category is Structure-based similarity: its aim is finding higher-level structures in the series to compare them on a more global scale.
 
Further, it is possible to subdivide this last category into two specific subcategories: Model-based distances work by fitting a model to the various series and then comparing the parameters of the underlying models. On the other hand, compression-based distances analyze how well two series can be compressed together. In this case, the similarity is reflected by higher compression ratios.

One fully extended method to compare time series is called Dynamic Time Warping. Dynamic time warping (DTW) is a well-known technique to find an optimal alignment between two given (time-dependent) sequences under certain restrictions. The sequences are warped in a nonlinear fashion to match each other. In fields such as data mining and information retrieval, DTW has been successfully applied to automatically cope with time deformations and different speeds associated with time-dependent data \cite{DBLPjournals/pr/PetitjeanKG11}.

Several algorithms exist in order to implement this technique and apply it to average signals  \cite{DBLPjournals/corr/Marteau15,ictdbid2292,Salvador,JSSv031i07}

Nevertheless, there is not a unique recipe that can be applied in the signal analysis. In some cases, there is the possibility to use a more heuristic approach, computationally cheap and also appropriate for data with regular time patterns. In this work, we propose a method that uses a signal envelope as a feature to align canary syllables and we provide the software implementation. 

The rest of the paper is organized as follows: In Section \ref{signal_aligment} the method is described in general terms with the details on each subsection. In Section \ref{code_implementation} a link to the software implementation is provided. Finally, in Section \ref{conclu} we present the conclusions and some further work.

\section{Signal alignment} \label{signal_aligment}

In present work we present an empirical algorithm that allows us to segment semi-periodic patterns and estimate the average of similar patterns. To reach this first is necessary to align the individual patterns and then estimate an average amplitude and time duration. Such information is of particular interest for research an characterization of birds singing in biology \cite{nature_bird_01}, thus method was applied to canary tonal sounds.

\begin{figure}[htb!]
\begin{center}
\hspace*{-0.5cm}\includegraphics[totalheight=5.2cm]{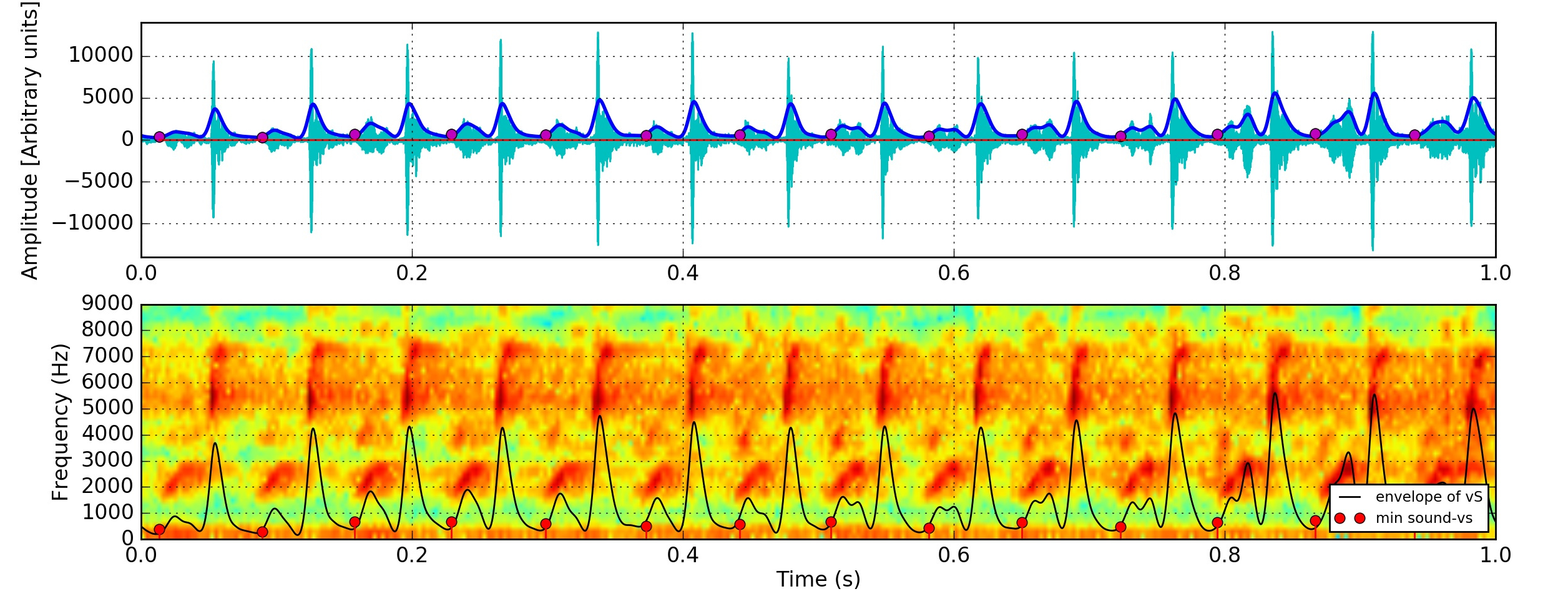}
\caption{An example of a canary song. The upper panel shows the amplitude signal vs.time. The bottom panel shows the sound frequency content as well as the envelope in arbitrary units of amplitude.}
\label{Fig_01}
\end{center}
\end{figure}

In data of this nature, different segments have differences in the time duration and amplitude, even when some features, like frequency content, remains the same. An example to describe a canary song is shown in Figure \ref{Fig_01}.

A canary can sing a repertory of different repetitions of syllables. Let us suppose that one wants to find a way to automatically measure the mean duration and amplitude of each kind of syllable. We describe here a way to automatically cut, align and estimate these mean quantities.

To summarize, the method consists of the following procedure: first, each sound segment is cut with a criterion that will be described in the Section \ref{signal_segment}. Second in order to align the many instances of the same syllable, once splitted, we calculate the envelope. Third, an anchor point is defined with a criterion that allows aligning signals with similar patterns. Finally, the signal length of each segment is re-sampled into a fixed length but preserving the individual duration of each segment in a vector. In the following subsections, we will describe each step of the process.

\subsection{Segmentation of the time series} \label{signal_segment}

\begin{figure}[htb!]
\begin{center}
\hspace*{0.1cm}\includegraphics[totalheight=4cm]{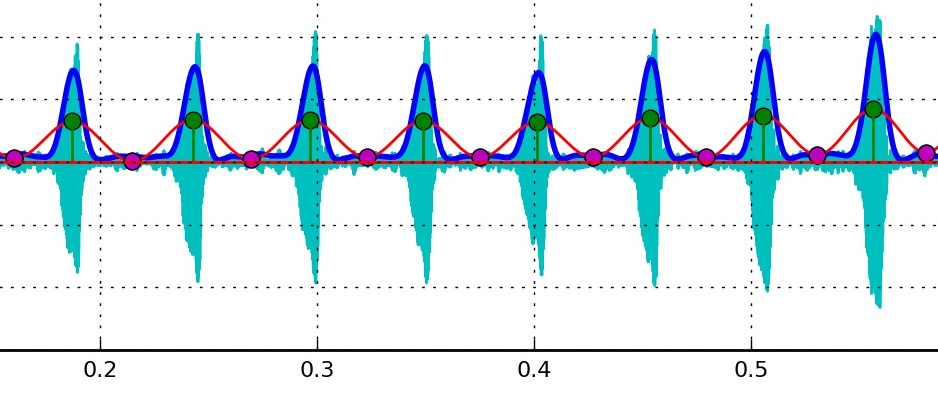}
\caption{An example of a canary song segmentation with the burly envelope in a pink line and the signal envelope in blue. }
\label{Fig_02}
\end{center}
\end{figure}

Each sound segment is cut automatically by means of the use of a burly envelope obtained with a low pass filter with cutoff frequency of the order of 20-40 Hz as shown in Figure \ref{Fig_02}. For a different data set the frequency of this filter is related to the frequency of the pattern repetition. The key to this step is to find the frequency repetition of the syllables and the minimum values to use them as an index to segment the series. 

That range was selected with respect to the canary syllable variations. We used the minimum of that envelope as an index to pre-cut each sound segment in individual syllables.
 
\subsection{Envelope estimation and anchor point} \label{anchor}

For each of those segments, the true envelope is estimated using the method detail described in \cite{paper_mio_envolvente}. During this procedure, the time duration is estimated for each segment and saved.

To synchronize all syllable of the multiple segments, we used as a feature the sound envelope. In some cases, we used the value of the relative maximum of the true envelope, in others the relative minimum and in some cases, the where the amplitude is variable or very fluctuating or the maximum is not well defined, we use the maximum of the low pass burly envelope as an anchor point. 

For each segment we include a margin of fix number of frames from the anchor point to preserve the entire shape of each pulse up to the signal start and end. In this way we used the feature anchor point to define a starting and ending point of each segment.

Regarding the selection of the feature, we ask the question of which is the better feature to define the anchor point? The answer is that it will depend on the particular characteristics of the signal. But we try the following procedure:

With the envelopes' amplitudes normalized to 1, we define candidates to anchor point as the earliest time $t_a$ in which the normalized amplitude has the value $a$ such as $0 < a \leq 1$.
We then displace all the envelopes from $t$ to $t' = t-t_a$; in this way, all of the syllables have the same normalized amplitude at $t'=0$. These are all the \emph{proposed} alignments.
For each proposed alignment, we calculate the mean squared error of the several envelopes, which is a measure of how well aligned the envelopes are. The anchor point is, then, the value that minimizes this mean squared error.

It is also worth noting that this procedure also gives a \emph{template} of the envelope of the syllables and, therefore, may be used to decide whether a syllable belongs to this group or not.

\begin{figure}[htb!]
\begin{center}
\hspace*{0.1cm}\includegraphics[totalheight=8.5cm]{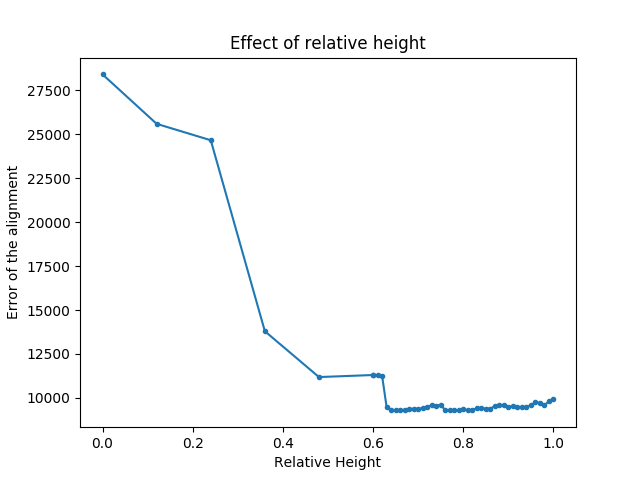}
\caption{A study of the mean squared error vs the selected anchor point. }
\label{Fig_02_b}
\end{center}
\end{figure}

\subsection{Signal re-sampling and averaging} \label{aver}

\begin{figure}[htb!]
\begin{center}
\includegraphics[totalheight=7.5cm]{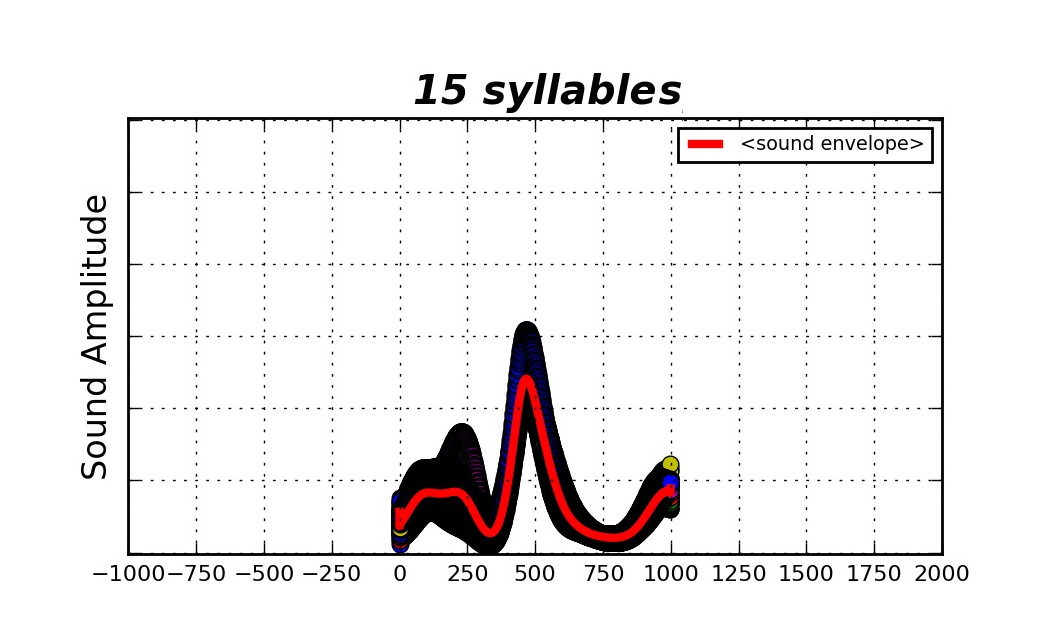}
\caption{The envelope of the song segments that have been re-sampled and belong to the Sound time series shown in Figure \ref{Fig_01}.}
\label{Fig_03}
\end{center}
\end{figure}

With the syllables aligned to this value, we average all the syllables. Using the anchor point, the mean value between all segment is obtained by first rescaling a re-sampling each segment to a fixed number of points (1000 points in the case of the example in Figure \ref{Fig_03}). In this way, every segment has a fixed number of points to be aligned.

\begin{figure}[htb!]
\begin{center}
\includegraphics[totalheight=7.5cm]{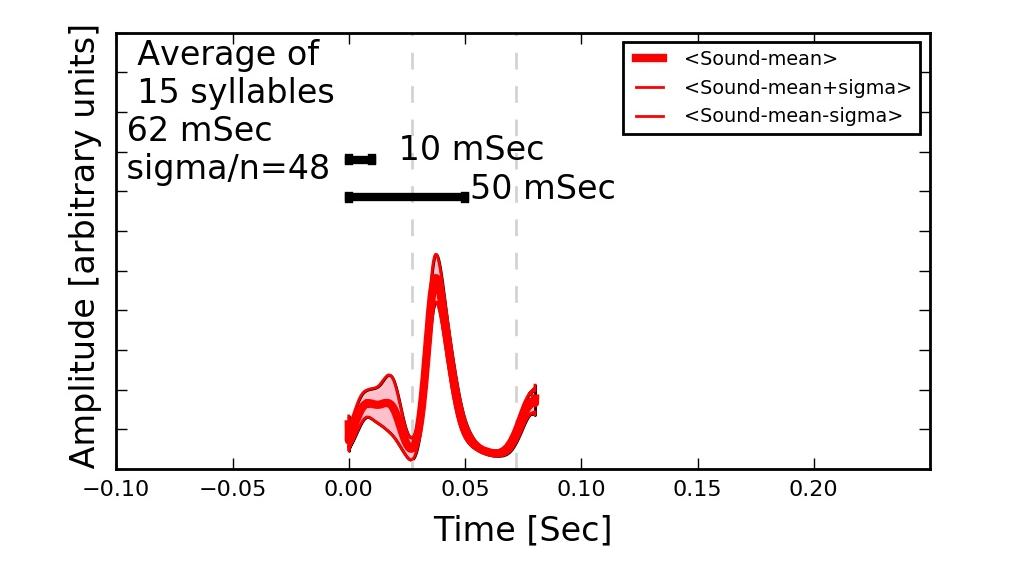}
\caption{Mean value of the envelope of the song segments. Time duration corresponds to the mean value of each segment time duration.}
\label{Fig_04}
\end{center}
\end{figure}

With all segments aligned, we take the average in each position. The result of this analysis is shown in Figure \ref{Fig_04}. If we want to have also a mean value represented with a mean duration we can use the average of each mean time duration of the segments \cite{1183920}.

In this way, when applying this procedure, we have an average syllable with an average amplitude, an average duration, and shape that could be useful for further segment classification.

\section{Regarding the code implementation} \label{code_implementation}

The software used for this method was developed on python. Open source scientific libraries where used such as Scipy and Numpy in order to be possible shear modify and improve the proposed algorithm. This procedure is Implemented in Python in $http://github.com/pabloalcain/syllable$


\section{Conclusions} \label{conclu}

In present work, a method to align and average segments or series with similar patterns is proposed. Even when the procedure is straightforward and simple, its a robust solution for different signals. An important advantage of this method is that it can be applied in signals with different spectral content. There are no requirements on the specific domain of the data. It is appropriate for one-dimensional data of any kind. Also, a procedure based on optimization is proposed for the anchor point selection, with the advantage that it could be adapted for different signal particularities.


\bibliography{mybibfile}
\bibliographystyle{spphys}       

\end{document}